\documentclass[12pt,reprint,amsmath,amssymb,aps,prd,twocolumn,superscriptaddress,floatfix,nofootinbib,longbibliography]{revtex4-2}

\usepackage{mathtools}
\usepackage{braket}

\usepackage{tikz}
\usepackage[colorlinks, allcolors=blue]{hyperref}
\usepackage{booktabs}
\usepackage{multirow}

\usepackage{csquotes}

\newcommand{\Z}{{\mathbb{Z}}}

\newcommand{\pqty}[1]{\left( #1 \right)}
\newcommand{\bqty}[1]{\left[ #1 \right]}
\newcommand{\abs}[1]{\left\lvert #1\right\rvert}

\newcommand{\expval}[1]{\langle #1 \rangle}

\newcommand {\dv}[3][ ]{
  \ifx #1 { }
    \frac{d #2}{d #3}
  \else
    \frac{d^{#1} #2}{d #3^{#1}}
  \fi
}

\newcommand {\pdv}[3][ ]{
  \ifx #1 { }
    \frac{\partial #2}{\partial #3}
  \else
    \frac{\partial^{#1} #2}{\partial #3^{#1}}
  \fi
}

\newcommand{\tr}{\operatorname{tr}}

\newcommand{\eps}{\epsilon}

\newcommand{\SU}{\mathrm{SU}}
\newcommand{\U}{\mathrm{U}}

\newcommand{\extParam}[0]{\alpha} %

\begin{abstract}

Inspired by self-adjoint extensions of the electric field operator in the Hamiltonian formalism, we extend the Wilsonian framework of Abelian lattice gauge theory by introducing a modified action parameterized by an angle $\extParam$, where the ordinary Wilson theory corresponds to $\extParam=0$. Choosing instead $\extParam=\pi$ (the ``staggered'' case) gives the only other theory in the family which preserves all symmetries of the original model at the microscopic level.
We study the case of $3D$ $\U(1)$ pure gauge theory, simulating the staggered case of this model numerically in its dual formulation. We find evidence of a continuum limit with a spontaneously broken $\Z_2$ single-site translational symmetry, in contrast to the ordinary theory. Moreover, the confining string fractionalizes into multiple strands which separate spatial regions in distinct ground states of the broken symmetry.

\end{abstract}

\begin{document}

\title{Broken Symmetry and Fractionalized Flux Strings \texorpdfstring{\\}{} in a Staggered U(1) Pure Gauge Theory}

\newcommand{\affilSaha}{\affiliation{Theory Division, Saha Institute of Nuclear Physics, 1/AF
Bidhan Nagar, Kolkata 700064, India}}
\newcommand{\affilHomiBhabha}{\affiliation{Homi Bhabha National Institute, Training School Complex,
Anushaktinagar, Mumbai 400094, India}}
\newcommand{\affilBern}{\affiliation{Albert Einstein Center for Fundamental Physics, Institute for Theoretical Physics,
University of Bern, Sidlerstrasse 5, CH-3012 Bern, Switzerland}}

\author{A. Banerjee}
\affilSaha
\author{D. Banerjee}
\affilSaha
\affilHomiBhabha
\author{G. Kanwar}
\affilBern
\author{A. Mariani}
\affilBern
\author{T. Rindlisbacher}
\affilBern
\author{U.-J. Wiese}
\affilBern

\maketitle

\newpage
 
\section{Introduction}

Symmetry is one of the most important organizing principles of quantum field theories. In the Wilsonian framework of renormalization, for example, one includes in the action of a quantum field theory all terms consistent with the symmetries of the model. In the context of lattice field theories, one expects that two actions which share a sufficiently large subgroup of spacetime, internal, and gauge
symmetries give rise to the same continuum theory. In lattice gauge theory, typical choices of the lattice action exactly implement gauge symmetry and the subgroup of Poincaré invariance associated with discrete lattice translations and hypercubic space-time rotations, but otherwise differ by irrelevant operators. Nevertheless, such choices of the action have been demonstrated to recover the continuum gauge theory physics in many theories.
For example, $\U(1)$ lattice gauge theory is commonly studied using either the standard Wilson action or the Villain action, which differ in their precise formulation but share the same symmetries and the same continuum limit~\cite{Villain:1974ir,Banks:1977cc,Creutz:1983ev}. 

From this point of view, it is natural to construct alternative discretizations for lattice gauge theories which may either recover the same continuum limit as standard theories or converge to a different continuum limit. In the former case, an alternative discretization can provide a better approach to the continuum, while in the latter case it can be useful from a model-building perspective.

In the present work we put forward a method of generating alternative Abelian lattice gauge theory actions inspired by first applying a \emph{self-adjoint extension} to the electric field operator in the Hamiltonian formalism, then establishing a path integral using Trotterization.
The use of self-adjoint extensions is natural in the operator language and manifestly preserves the operator commutation relations that define the theory.
On the other hand, the resulting lattice gauge theories are far from obvious from a Euclidean action perspective.
In this sense, our work expands the current framework of lattice gauge theories, potentially enabling other such reformulations of gauge theories in ways that are either beneficial for simulation or yield new continuum gauge theories.

As a demonstration of the approach, we formulate a class of $\U(1)$ lattice gauge theories parameterized by an angle $\extParam \in [0, 2\pi)$. The choice $\extParam = 0$ corresponds to the Villain action, while other choices correspond to novel lattice actions with $\U(1)$ gauge symmetry. Taking $\extParam = \pi$ is particularly interesting as it is the only other choice which preserves all relevant symmetries of the theory, including charge conjugation and parity. In this work, we therefore focus on numerically studying the continuum limit of the $\extParam = \pi$ theory in three spacetime dimensions. In this case, the symmetry group of lattice translations is partially broken, as translations by a single site in any direction remain a symmetry of the theory only if combined with an appropriate internal symmetry transformation, which we together denote as a ``single-site shift'' symmetry. This motivates us to call this the ``staggered'' case. Translations by an even number of lattice spacings remain unbroken, which we expect to be sufficient,  together with the hypercubic space-time symmetry, to recover full Poincar\'{e} invariance in the continuum.

Numerical simulation demonstrates that the single-site shift symmetry is spontaneously broken and remains so as the continuum limit is approached. These simulations also demonstrate that the confining string of the theory fractionalizes into two strands separating inner and outer regions of space which exist in different vacua of the spontaneously broken symmetry. These observations suggest a continuum gauge theory that is distinct from the usual $\U(1)$ pure gauge theory in 3D. Our numerical conclusions are supported by an effective theory for the $\extParam=\pi$ theory which is derived analytically. 

Several features of this model, including a phase diagram characterized by the breaking of single-site shifts and charge conjugation, as well as flux string fractionalization, have been previously observed in the context of quantum link models which display crystalline and nematic confined phases \cite{QuantumLink, QuantumDimer1, QuantumDimer2, NematicTriangular}, and the former is also observed in the $\SU(N)$ quantum spin ladder regularization of 2D $\mathrm{C}\mathrm{P}(N-1)$ models at $\theta = \pi$ \cite{QuantumSpinLadder1, QuantumSpinLadder2}.
To the best of our knowledge, this work represents the first time where such phenomena are observed in a lattice gauge theory with an infinite-dimensional Hilbert space.  Moreover, systems where single-site shifts play an important role have gained interest in recent years in the context of ``emanant'' symmetries \cite{Emanant1,Emanant2}.

The remainder of this work is organized as follows. In Section~\ref{sec:usual u(1)}, we review analytical and numerical results for the standard three-dimensional $\U(1)$ lattice gauge theory. In Section~\ref{sec:new u(1)} we introduce the self-adjoint extension of this $\U(1)$ gauge theory and explain how the $\extParam$ parameter is introduced in the case of general dimensionality.
We then specialize to three dimensions and consider the staggered (i.e., $\extParam=\pi$) theory, discussing its dualization to a height model which allows numerical simulation without a sign problem.
Section~\ref{sec:numerical} details the numerical investigations of the dual model.
We investigate order parameters for the breaking of the relevant symmetries, and we find evidence that the $\extParam=\pi$ theory has a broken $\Z_2$ symmetry down to the continuum limit, a feature which is absent from the standard $\U(1)$ theory. To further characterize the $\extParam=\pi$ theory, we also present numerical calculations of its mass and string tension. In Section \ref{sec:effective theory}
we analytically derive an effective theory from the microscopic degrees of freedom, which is consistent with the results obtained via numerical simulation. Finally, in Section \ref{sec:conclusions} we summarize and conclude with an outlook on future steps.

\section{U(1) lattice gauge theory in 3D}
\label{sec:usual u(1)}

In this section we review the standard $\U(1)$ lattice gauge theory in three dimensions, as it is understood analytically and numerically in both the action and Hamiltonian formulation. While our construction of the $\extParam \neq 0$ Abelian gauge theory is valid in any dimension, as we will see in later sections, in this work we focus on the three-dimensional case. As such, an understanding of the standard three-dimensional Abelian gauge theory sets a baseline expectation against which the $\extParam \neq 0$ theory can be compared.

\subsection{Action formulation of the standard Abelian gauge theory}

The standard path integral formulation of $\U(1)$ lattice gauge theory in 3D is defined in terms of $\U(1)$-valued variables arranged on the links of a 3D Euclidean spacetime lattice. The discretized action is commonly chosen to be one of two equivalent formulations, either the Wilson action or the Villain action. The partition function in the Villain formulation is given by
\begin{multline}
    \label{eq:usual u(1) partition function}
    Z = \pqty{\prod_{l \in \mathrm{links}} \int_{0}^{2\pi} d\varphi_l} \pqty{\prod_{p \in \mathrm{plaq}} \sum_{n_p = -\infty}^{+\infty}} \\
    \times  \exp{\bqty{ -\frac{1}{2e^2}\sum_p ((d\varphi)_p - 2\pi n_p)^2}} \ ,
\end{multline}
where $l$ runs over links connecting neighboring sites of the lattice and $p$ runs over the plaquettes of the lattice. Labelling the ordered links in the plaquette $p$ from $1$ to $4$, we define
\begin{equation} \label{eq:plaquette}
(d\varphi)_p \equiv \varphi_1 + \varphi_2 - \varphi_3 - \varphi_4 \ .
\end{equation}
In Eq.~\eqref{eq:usual u(1) partition function} we have absorbed the lattice spacing $a$ in the dimensionless coupling $e^2$. Unless otherwise specified, in the rest of this work we set $a=1$.

The theory has been rigorously shown to be confining at all values of the coupling and the asymptotic scaling of the mass and string tension near the continuum limit $e^2 \to 0$ have also been analytically computed~\cite{GopfMack}. In particular, the mass gap $m$ scales as (restoring the lattice spacing $a$) %
\begin{equation}
    \label{eq:mass gap}
    a^2 m^2 = \frac{8\pi^2}{e^2} \exp{\bqty{-2\pi^2 v_0/e^2}} \ ,
\end{equation}
where $v_0 \approx 0.2527$.
On the other hand, the string tension $\sigma$ scales as (again, restoring the lattice spacing $a$)
\begin{equation}
    \label{eq:string tension analytical}
    a^2 \sigma = \frac{\widetilde{c}}{4\pi^2}\, a m \,e^2  \ ,
\end{equation}
in terms of a dimensionless constant $\widetilde{c}$.
These results have found numerical confirmation~\cite{TepAthen, CasPan, Loan2003, Loan:2003wy}, and the constant $\widetilde{c}/4\pi^2 \approx 0.21$ was estimated in \cite{TepAthen}. As the continuum is approached, we then see that the dimensionless ratios
\begin{equation}
    \label{eq:ratio scaling}
    \frac{\sigma}{m^2} \to \infty  \ , \quad \quad \quad \frac{(a^2 \sigma)}{(am)\, e^2} \to \frac{\widetilde{c}}{4\pi^2} \ .
\end{equation}
Equation~\eqref{eq:ratio scaling} shows that the theory has several inequivalent length scales, even in the continuum. As such, different continuum limits may be obtained depending on what quantity is taken as the standard of length. In a continuum limit where the mass $m$ is held fixed in physical units,\footnote{In this case, Eq.~\eqref{eq:mass gap} implicitly defines a function $e^2(a)$ which represents the renormalization trajectory of the dimensionless coupling as the continuum limit is approached while $m$ is kept fixed, such that $e^2(a) \to 0$ as $a\to 0$ (and therefore also $am \to 0$ since $m$ is held fixed).} the string tension becomes infinite, and the theory is equivalent to a free scalar (the ``photoball'') of mass $m$ \cite{GopfMack}.
If the string tension $\sigma$ is held fixed instead, the mass (in physical units) goes to zero. Finally, the dimensionful bare coupling $e^2/a$ has dimensions of energy, and can be held fixed as a third possible prescription to obtain a continuum limit, with the corresponding continuum theory conjectured to be free electrodynamics~\cite{GopfMack}. For further discussion of the continuum limits of this theory, see \cite{TepAthen}.

The analytical and numerical analysis of the theory is much simplified in its dual formulation. In particular, the partition function in Eq.~\eqref{eq:usual u(1) partition function} can be rewritten in terms of integer-valued height variables $h_x \in \Z$ associated with sites $x$ of the dual lattice. In terms of these new variables, the partition function is given (up to constant prefactors) by~\cite{Banks:1977cc,Savit:1977fw} %
\begin{equation}
    \label{eq:unstaggered height model}
    Z = \pqty{\prod_{x \in \mathrm{sites}} \sum_{h_x=-\infty}^{+\infty}}\exp{\bqty{-\frac{e^2}{2} \sum_{\expval{xy}} (h_x-h_y)^2}} \ ,
\end{equation}
where the product over $x$ enumerates sites of the dual lattice and the sum over $\expval{xy}$ enumerates pairs of neighboring dual sites $x$ and $y$.

The dualization forms the basis of our analytical understanding of the theory near the continuum. Near the continuum limit the integer height variables can be replaced with real scalars, which results in the Sine-Gordon model~\cite{GopfMack, SemiAbelian}
\begin{multline}
    \label{eq:usual u(1) effective theory}
    Z =\pqty{\prod_x \int_{-\infty}^{+\infty} d\phi_x }\exp\bigg[-\frac{1}{2} \sum_{\expval{xy}} (\phi_x-\phi_y)^2 +\\
    +2e^{-2\pi^2 v_0 /e^2}\sum_x \cos{\pqty{\frac{2\pi\phi_x}{\sqrt{e^2}}}} \bigg] \ ,
\end{multline}
where $\phi_x$ is a real scalar field. This provides an effective description of the theory that is valid for small $e^2$. In Section \ref{sec:effective theory} we extend these results to the $\extParam=\pi$ case.

\subsection{Hamiltonian formulation of the standard Abelian gauge theory}

As we will see in the next sections, the $\extParam \neq 0$ $\U(1)$ gauge theory is most easily understood in the Hamiltonian formulation. In this section, we review the Hamiltonian formulation of the standard $\U(1)$ gauge theory.

In the Hamiltonian formulation of lattice gauge theories \cite{KogSuss}, time is continuous while space is discretized into a square (in general: hypercubic) lattice. The temporal gauge $A_0=0$ is chosen. A classical configuration of the theory is given by an assignment of a group-valued variable $U_l \in \U(1)$ to each spatial link $l$ of the lattice. In the quantum theory, the Hilbert space on each link $l$ is therefore given by $\mathcal{H}_l \equiv L^2(\U(1))$, the space of square-integrable functions on $\U(1)$. The elements $\ket{\psi}$ of this space may be expanded in terms of wavefunctions $\psi(\varphi_l)$ over the angular variable $\varphi_l \in [0, 2\pi]$ associated with the $\U(1)$ group element by $U_l = \exp{\pqty{i\varphi_l}}$ as
\begin{equation}
    \ket{\psi} = \int_0^{2\pi} d\varphi_l\, \psi(\varphi_l) \ket{U_l} \ ,
\end{equation}
where the orthonormal basis $\{\ket{U}\}$ may be interpreted as a position basis in group space. The wavefunctions satisfy
\begin{equation}
    \label{eq:periodic bcs}
    \psi(2\pi)=\psi(0), \quad \quad \int_0^{2\pi} d\varphi \abs{\psi(\varphi)}^2 < \infty \ .
\end{equation}
The total Hilbert space $\mathcal{H}_{\mathrm{tot}}$ is then given by the tensor product of these spaces over all links,
\begin{equation}
    \mathcal{H}_{\mathrm{tot}} = \bigotimes_{l \in \mathrm{links}} \mathcal{H}_l \ . %
\end{equation}

The Hamiltonian may be obtained by studying the transfer-matrix formulation of the path integral and taking the continuous-time limit, giving 
\begin{equation}
    \label{eq:kogut susskind hamiltonian}
    H = \frac{e^2}{2} \sum_{l \in \mathrm{links}} E_l^2 + \frac{1}{2e^2} \sum_{p \in \mathrm{plaqs}} B^2((d\varphi)_p) \ .
\end{equation}
Here $(d\varphi)_p$ is the plaquette variable on spatial plaquettes $p$ defined as in Eq.~\eqref{eq:plaquette} above, while the electric field operator $E_l$ on spatial link $l$ is given by 
\begin{equation}
    E_l = -i\pdv{}{\varphi_l}  \ .
\end{equation}
Therefore, on each link the theory is analogous to the familiar problem of a quantum mechanical particle on the circle \cite{ParticleCircle, Tong}: the $\U(1)$ variable $U_l = \exp{(i\varphi_l)}$ identifies the position of the particle on the circle, while the electric field $E_l$ is the canonical momentum conjugate to the position. The magnetic energy term $B^2(\cdot)$ couples these links together, and for the standard $\U(1)$ gauge theory it may be chosen as the Wilson term, Villain term, or any other term equivalent up to lattice artifacts. In the previous section, we have chosen the Villain formulation for the Euclidean action of the theory, since it leads to the simpler partition function. Therefore, we here also choose the Villain term, which is easiest to define in exponential form as
\begin{multline}
    \label{eq:villain magnetic term standard}
    \exp\pqty{-B^2((d\varphi)_p)/2e^2} \\
    = \sum_{n \in \Z} \exp{\pqty{-\frac{1}{2e^2} \sum_p ((d\varphi)_p - 2\pi n)^2}} \ .
\end{multline}
For comparison, the Wilson plaquette term would take the form $B^2((d\varphi)_p) = 1-\cos{((d\varphi)_p)}$.

In order to correctly implement the gauge symmetry, only some states in the total Hilbert space $\mathcal{H}_{\mathrm{tot}}$ should be considered as physical. In particular, one includes in the physical Hilbert space $\mathcal{H}_{\mathrm{phys}}$ only those states $\ket{\psi}$ which satisfy the Gauss law constraint
\begin{equation}
    \label{eq:gauss law}
    G_x \ket{\psi} = 0 \ , \quad\quad G_x = \sum_{i} \pqty{E_{x,x+\hat{i}}-E_{x-\hat{i},x}} \ ,
\end{equation}
where the sum over $i$ runs over all spatial directions and $\hat{i}$ is the unit vector oriented in the $i$-th spatial direction. 
Both choices of magnetic energy $B^2(\cdot)$ discussed above are designed to commute with the Gauss law operators $G_x$, so that the $G_x$ operators commute with the Hamiltonian and gauge symmetry is properly respected. 

\section{U(1) gauge theory with an \texorpdfstring{$\extParam$}{alpha} angle}\label{sec:new u(1)}

We next describe the construction of a class of Abelian gauge theories inspired by self-adjoint extensions. Starting from the Hamiltonian formulation of the standard $\U(1)$ gauge theory, we extend its Hilbert space in the most general way consistent with the gauge symmetry. The possible extensions are characterized by an angle $\extParam$, where the choice $\extParam = 0$ corresponds to the standard $\U(1)$ theory. The action formulation of the theory is then constructed via Trotterization from the Hamiltonian. In order to preserve cubic symmetries of the action, and thus Lorentz invariance in the continuum limit, the magnetic terms in the action are appropriately modified. The resulting theory explicitly breaks charge conjugation and parity unless $\extParam = 0$ or $\extParam = \pi$. We therefore choose to focus on $\extParam = \pi$ for more detailed study.
This construction, which is quite natural in the Hamiltonian formulation, leads to a fairly complicated action which does not fall within the Wilsonian framework of gauge theories, but still shares all the symmetries of the standard $\U(1)$ gauge theory. Importantly, this includes exact $\U(1)$ gauge symmetry, which is therefore inherited by the continuum theory. While the construction applies to arbitrary spacetime dimension, we further focus on the three-dimensional case and dualize the theory, which removes the sign problem present in the original action formulation and results in a theory suitable for numerical study.

\subsection{Self-adjoint extension of the Hamiltonian}

We have seen in the previous sections that the Hilbert space of the standard $\U(1)$ gauge theory is given, on each lattice link, by the square-integrable functions on $\U(1)$. In the usual formulation, the state $\ket{\psi}$ is defined by wavefunctions satisfying the periodicity and integrability conditions in Eq.~\eqref{eq:periodic bcs}.
The left side of Eq.~\eqref{eq:periodic bcs} ensures that the wavefunctions are continuous on the periodic $\U(1)$ manifold. However, the probabilistic interpretation of quantum theories only requires that the absolute value squared $\abs{\psi(\varphi)}^2$ be periodic on $\U(1)$ for the wavefunction to provide a representation of the $\U(1)$ symmetry. It is therefore consistent to relax the first condition of Eq.~\eqref{eq:periodic bcs} and instead require only that the wavefunction be periodic up to an arbitrary phase,
\begin{equation}
    \label{eq:twisted periodic bcs}
    \psi(2\pi)=e^{i\extParam}\psi(0) \ .
\end{equation}
Such a modification is familiar from the quantum mechanics of a rotor (see for example \cite{ParticleCircle}). In fact, on each link the electric Hamiltonian $-\frac{e^2}{2} \partial_\varphi^2$ is equivalent to the Hamiltonian of a free non-relativistic particle on the circle. Considering this fictitious particle to have charge $1$, twisting the wavefunctions by the $\extParam$ angle is equivalent to threading a magnetic flux equal to $\extParam$ through the circle on which this charged particle lives.

The choice in Eq.~\eqref{eq:twisted periodic bcs} may also be understood in a more mathematical way as a \textit{self-adjoint extension} \cite{ReedSimon, Gieres}. The Hamiltonian of the standard $\U(1)$ gauge theory involves the electric field operator $E_l = -i\partial/\partial \varphi_l$, which acts separately on the Hilbert space of each link. A basic requirement is that the electric field must be \textit{self-adjoint}, so that it has real eigenvalues and an orthonormal basis of eigenfunctions. This is necessary so that the Hamiltonian and the Gauss law have their expected properties. Equation~\eqref{eq:twisted periodic bcs} then describes the most general choice for the wavefunction $\psi(\varphi)$ which is consistent with self-adjointness of the electric field.

In principle, this choice only modifies the Hilbert space of the theory, and operators such as the electric field and the Hamiltonian are only affected through their domain of definition. For simplicity, however, we choose to
map the Hilbert space of twisted wavefunctions defined by Eq.~\eqref{eq:twisted periodic bcs} to the ordinary Hilbert space of periodic wavefunctions
via
\begin{equation}
    \label{eq:mapping twisted to periodic}
    \psi(\varphi) \rightarrow e^{-i \varphi \extParam / 2 \pi} \psi(\varphi) \ .
\end{equation}
This mapping modifies the definition of the electric field operator, sending it to
\begin{equation}
    \label{eq:modified electric field}
    E_l' = E_l +\tfrac{\extParam}{2\pi} \ .
\end{equation}
The family of theories defined by the introduction of the angle $\extParam$ can then by identified with theories over
the ordinary Hilbert space, where instead the Hamiltonian is defined in terms of a modified electric field as in Eq.~\eqref{eq:modified electric field}.
The family of Hamiltonians obtained by all such modifications of the electric field operator are naturally motivated by self-adjoint extensions and maintain the gauge symmetry of the theory.
Defining the self-adjoint extension by Eq.~\eqref{eq:modified electric field} also makes it clear that this is still consistent with the Gauss law constraint given in Eq.~\eqref{eq:gauss law}.

The choice of wavefunctions in Eq.~\eqref{eq:twisted periodic bcs}, or the equivalent modification to the electric field operator in Eq.~\eqref{eq:modified electric field}, generally affects the other symmetries of the theory. Under charge conjugation, the link variables transform as $U_l \to U_l^*$ and the electric fields transform as $E_l \rightarrow -E_l$. This maps $\extParam \rightarrow -\extParam$ in either the wavefunction definition or electric field definition. Therefore unless $\extParam \in \{0,\pi\}$, the resulting theory explicitly breaks charge conjugation.
There is also an implicit choice of orientation in the definition of $\extParam$, which is more clearly seen when considering the modification to the electric field operator in Eq.~\eqref{eq:modified electric field}, because this operator is directional. Choosing $\extParam \in \{0,\pi\}$ also ensures that the theory preserves parity, which reverses this direction and thus sends $\extParam \rightarrow -\extParam$. In fact, discrete rotations would also be broken by a choice of $\extParam \not\in \{0,\pi\}$ unless the orientations of these terms are carefully chosen across all links, further motivating the restriction to these values.

In $(1+1)$D, the substitution \eqref{eq:modified electric field} in the gauge theory Hamiltonian is equivalent to introducing a topological $\theta$ term where $\theta$ corresponds to $\extParam$ \cite{Tong}. In $(3+1)$D, the topological $\theta$ term is instead introduced by the replacement $\vec{E} \to \vec{E} -\tfrac{\theta}{8\pi^2} \vec{B}$, which preserves Lorentz-invariance \cite{Tong}. On the other hand, the replacement \eqref{eq:modified electric field} is \textit{not} Lorentz-invariant in dimensions higher than $(1+1)$D, and, as we will see in a moment, we therefore need to appropriately modify the partition function of the theory in order to restore Lorentz invariance. In this sense the $\extParam$ angle represents one possible higher-dimensional generalization of the two-dimensional $\theta$ term.

\subsection{Action formulation for the \texorpdfstring{$\extParam \neq 0$}{} theory}

The action formulation can be obtained from the Hamiltonian formulation via Trotterization. In the \enquote{position basis} $\{ \ket{U} \}$ for $U = e^{i \varphi} \in \U(1)$, the magnetic Hamiltonian is diagonal. The matrix elements of the exponential of the electric Hamiltonian can be computed by inserting a basis of \enquote{momentum eigenstates} $\{\ket{m}\}$, where $m \in \Z$, which satisfy
\begin{equation}
    \expval{U | m} = \frac{1}{\sqrt{2\pi}} e^{im\varphi}
\end{equation}
and diagonalize the electric field $E=-i \partial_{\varphi} +\tfrac{\extParam}{2\pi}$ according to
\begin{equation}
    E \ket{m} = \pqty{m + \frac{\extParam}{2\pi}} \ket{m} \ .
\end{equation}
By inserting resolutions of the identity in terms of $\{ \ket{m} \}$ and applying Poisson summation, one obtains 
\begin{multline}
    \label{eq:electric matrix elements}
    \bra{U'} e^{-\Delta \tau \frac{e^2}{2} E^2} \ket{U}
    = \frac{1}{\sqrt{ 2\pi e^2 \Delta \tau}} e^{-i\tfrac{\extParam}{2\pi}(\varphi'-\varphi) } \times \\ \times \sum_{m \in Z} \exp{\pqty{-\frac{1}{2 e^2 \Delta \tau} \pqty{\varphi'-\varphi - 2\pi m }^2 }} e^{i \extParam m} \ .
\end{multline}
Since we are working in the temporal gauge, the matrix element above describes the path integral weight of each timelike plaquette.
Meanwhile, the path integral weight of each spacelike plaquette is simply given by the exponential $e^{-\Delta \tau B^2((d\varphi)_p)}$, because the operator $(d\varphi)_p$ is diagonal in the ``position basis'' $\{ \ket{U} \}$ describing the link variables in the path integral.

As remarked in the Introduction, we aim to construct a theory which shares as many of the symmetries of the standard $\U(1)$ theory as possible. In particular, we would like to construct a theory which is invariant under the hypercubic subgroup of the Lorentz symmetry that remains on the Euclidean lattice.
Since we are interested in working on isotropic lattices, one can set as usual $\Delta \tau = 1$. For $\extParam=0$, the weight of the timelike plaquettes given in Eq.~\eqref{eq:electric matrix elements} is then identical to the weight of the spacelike plaquettes using the Villain term given in Eq.~\eqref{eq:villain magnetic term standard}.
This leads to the partition function in Eq.~\eqref{eq:usual u(1) partition function} for the usual theory with the Villain action, which is manifestly isotropic and invariant under the group of Euclidean hypercubic symmetries.

In order to obtain an isotropic lattice theory for $\extParam \neq 0$, we match the spatial plaquette terms in the lattice action to the space-time plaquette terms Eq.~\eqref{eq:electric matrix elements} derived by the Trotterization steps above. We note that the Hamiltonian arising from considering the one-timestep transfer matrix then includes an imaginary magnetic field term $B^2(\cdot)$. However the two-timestep transfer matrix remains positive definite and gives rise to a Hermitian Hamiltonian, indicating that this still corresponds to a well-defined quantum theory.
The partition function for generic $\extParam$ is then given by 
\newcommand{\extPartFn}[0]{
Z = \pqty{\prod_{l} \int_0^{2\pi} d\varphi_l } \prod_p \Big[ \sum_{m \in \mathbb{Z}} e^{-i\tfrac{\extParam}{2\pi} (d\varphi)_p } \times \\ \times \exp{\pqty{-\frac{1}{2 e^2 } \pqty{(d\varphi)_p - 2\pi m }^2 }} e^{i \extParam m} \Big]
} %
\begin{multline} \label{eq:extended partition fn}
\extPartFn \ .
\end{multline}
This partition function is gauge invariant and is invariant under cubic rotations. For $\extParam=0$ the partition function Eq.~\eqref{eq:extended partition fn} reduces to the partition function Eq.~\eqref{eq:usual u(1) partition function} of the ordinary theory. As remarked above, choosing $\extParam = 0$ or $\extParam = \pi$ also explicitly preserves parity invariance of the action, making the action invariant under the full cubic symmetry subgroup of the Lorentz symmetry.

\begin{figure}
    \centering
    \includegraphics[width=3cm]{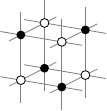}
    \caption{A subset of the three-dimensional dual lattice, depicting the staggered integer (black dots) and half-integer (white dots) height variables associated respectively with even and odd dual sites.}
    \label{fig:staggered lattice}
\end{figure}

While the partition function in Eq.~\eqref{eq:extended partition fn} generally suffers from a sign problem, this can be completely resolved by dualization.
Following a similar procedure to the one used to arrive at Eq.~\eqref{eq:unstaggered height model} in the $\extParam = 0$ theory, the partition function for the three-dimensional $\extParam = \pi$ theory can be dualized to that of a height model,
\begin{multline}
    \label{eq:staggered height model}
    Z = \pqty{\prod_{x\, \mathrm{odd}} \sum_{h_x \in \Z + \frac{1}{2}}} \pqty{\prod_{y\, \mathrm{even}} \sum_{h_y \in \Z}} \times \\ \times \exp{\bqty{-\frac{e^2}{2} \sum_{\expval{xy}} (h_x-h_y)^2}} \ .
\end{multline}
Details of the dualization are presented in Appendix~\ref{sec:dualization}.
Here each lattice site at position $\vec x = (x_0, x_1, x_2)$ is said to be even or odd according to the parity of $x_0+x_1+x_2$, as shown in Fig.~\ref{fig:staggered lattice}.
The only difference compared to Eq.~\eqref{eq:unstaggered height model} is the assignment of half-integer height variables on the odd sites of the lattice.
In particular, all the nearest neighbours of integer-valued height variables are half-integer, and vice versa. This should be contrasted with the related case of quantum link models, where the staggering is only realized in the space directions \cite{QuantumLink,QuantumDimer1,QuantumDimer2,NematicTriangular}.

\subsection{Symmetries and order parameters}\label{sec:symmetries}

The staggered height model defined by the partition function Eq.~\eqref{eq:staggered height model} enjoys the following symmetries:
\begin{enumerate}
    \item \textit{Global $\Z$-invariance}: $h_x \to h_x + c$ where $c$ is any constant integer. Note that under this transformation the integer and half-integer nature of the height variables is preserved. Importantly, this symmetry should be understood as a redundancy in our description of the system, whereby the overall height around which the height variables fluctuate is irrelevant \cite{GopfMack}. All observables are therefore required to be $\Z$-invariant. 
    \item \textit{$\Z_2$ charge conjugation $C$}: $h_x \to -h_x$. Again, this transformation respectively maps integers and half-integers to integers and half-integers. Note that the standard $\U(1)$ theory also enjoys this symmetry, but for $\extParam \neq 0$, only the choice $\extParam=\pi$ leads to an action invariant under charge conjugation. 
    \item \textit{$\Z_2$ single-site shift symmetry $S$}: $h_x \to h_{x+\hat{\mu}}+\frac12$ where $\mu$ is any spacetime direction. Note that the half-integer offset is required in order to preserve the integer and half-integer nature of the height variables.
    While $S$ as defined is not technically speaking a $\Z_2$ symmetry (i.e.\ it does not square to the identity, but rather to a translation), it can be made so by combining it with parity $P$ and charge conjugation $C$ in any order. For example $CPS: h_x \to -h_{-x+\hat{\mu}}-\frac12$ squares to the identity.   
    \item \textit{Translations by an even number of lattice spacings}: While translations by one lattice spacing swap integers and half-integers, translations by an even number of lattice spacings preserve the nature of the height variables. We expect that invariance under these symmetries is enough to recover full translational invariance in the continuum.
    \item \textit{Remnant Lorentz symmetry}: The action of the height model is fully isotropic and thus invariant under the group of Euclidean cubic rotations and, for $\extParam=\pi$, also reflections. This corresponds to all spacetime symmetries of the cubic lattice, and
    we therefore expect to recover a Lorentz-invariant theory in the continuum.
\end{enumerate}
An analogy for the shift symmetry
comes from staggered fermions, where the fermion degrees of freedom are spread over multiple lattice sites and translations by one lattice site are a symmetry of the action only when combined with an extra internal rotation \cite{StaggeredSymmetry}. The single-site shift symmetry of staggered fermions can break spontaneously \cite{StaggeredBreaking}, and, as we will see, this also happens in our model. Moreover, quantum link models \cite{QuantumLink,QuantumDimer1,QuantumDimer2,NematicTriangular}, as well as the quantum spin ladder regularization of $\mathrm{C}\mathrm{P}(N-1)$ models \cite{QuantumSpinLadder1, QuantumSpinLadder2}, provide further examples of theories with a phase diagram characterized by the breaking of charge conjugation and single-site shifts.  

In order to investigate the possible breaking of charge conjugation $C$ and the shift symmetry $S$, we construct appropriate order parameters, along the lines of \cite{QuantumLink,QuantumDimer1,QuantumDimer2,NematicTriangular,QuantumSpinLadder1, QuantumSpinLadder2}. One order parameter is defined by
\begin{equation}
    O_{CS} = \sum_{x} (-1)^x h_x = \sum_{x\,\mathrm{even}} h_x - \sum_{x\,\mathrm{odd}} h_x \ ,
\end{equation}
and changes sign under either $S$ or $C$. Note that $O_{CS}$ is a sum of local observables and invariant under the global $\Z$-symmetry. We expect $O_{CS}$ to acquire a vacuum expectation value only if both $C$ and $S$ are spontaneously broken; either symmetry remaining unbroken is sufficient for $O_{CS}$ to have zero vacuum expectation value. It is therefore important to also construct an observable which is sensitive to only one of the two symmetries. In particular, we define a second order parameter
\begin{equation}
    O_S = \sum_{c \in \mathrm{cubes}} \sum_{x \in c} (-1)^x (h_x - \bar{h}_c)^2 \ ,
\end{equation}
where the sum runs first over all elementary cubes $c$ in the dual lattice, and then over dual sites $x$ within each cube. The average height variable within the cube is defined as
\begin{equation}
    \bar{h}_c = \frac{1}{8} \sum_{x \in c} h_x \ .
\end{equation}
The observable $O_S$ is $C$-invariant, but changes sign under the shift symmetry $S$. We therefore expect it to acquire a vacuum expectation value if the single-site shift symmetry $S$ is broken. The somewhat complex construction of $O_S$ is required in order to obtain an observable with the correct symmetry properties, which is at the same time the sum of local terms (the cubes) and invariant under the global $\Z$ symmetry.

\section{Numerical simulation}\label{sec:numerical}

In order to investigate the $C$ and $S$ symmetry structure and the phase diagram of the theory, we numerically simulated the staggered height model using cluster Monte Carlo algorithms \cite{Evertz,wolff,wang-swendsen}.
The simulations were performed on $L^3$ lattices from $L=32$ up to $L=256$ and couplings between $e^2 = 0.3$ and $e^2 = 2.0$. As remarked in previous sections, we expect the continuum limit as $e^2 \to 0$, much like in the standard $\U(1)$ gauge theory. Unless specifically noted, all quantities are given in lattice units in the following discussion.

\subsection{Order parameters}\label{sec:order parameters}

\begin{figure*}
    \centering
    \includegraphics{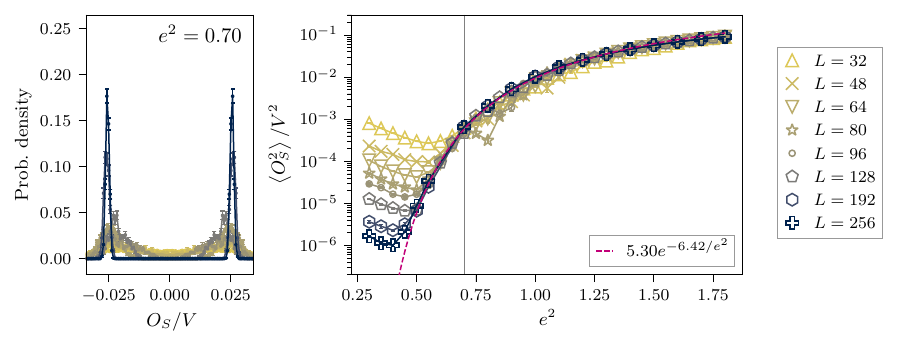}
    \caption{Left: Histogram of the operator $O_S$ normalized by the volume $V$ at $e^2=0.70$. Right: Susceptibility of $O_S$ normalized by $V^2$ as a function of $e^2$ for several volumes, with a fit of the large-volume data to the form $A \exp{(-B / e^2)}$. The vertical line is located at the coupling $e^2=0.70$ where the histogram is shown.  The double-peaked structure at large volume (left) and scaling as $V^2$ (right) indicates spontaneous breaking of the $S$ symmetry.}
    \label{fig:OS observable data}
\end{figure*}

To study the symmetry structure of the theory, it is useful to consider histograms of the order parameters $O_S$ and $O_{CS}$ as well as the normalized susceptibilities $O_S^2 / V^2$ and $O_{CS}^2 / V^2$ in terms of the lattice volume $V = L^3$. For a spontaneously broken symmetry, we expect to see a volume-independent susceptibility and a double-peaked histogram for sufficiently large volumes. This implies that the relevant operator acquires a vacuum expectation value.

The operator $O_S$ demonstrates a clear signal of spontaneous breaking of $S$ symmetry over the couplings studied. The relevant numerical data is shown in Fig.~\ref{fig:OS observable data}. In particular, for couplings $e^2 \gtrsim 0.65$ the normalized susceptibility $O_S^2/V^2$ is essentially volume-independent. Further confirmation can be found in the histogram of the values of $O_S/V$, which shows two clearly defined peaks becoming sharper as the volume is increased.
For $e^2 \lesssim 0.65$, the normalized susceptibility $O_S^2/V^2$ begins to develop volume dependence, with decreasing susceptibility as the volume increases until the volume is sufficiently large. This indicates that many of the ensembles correspond to physical volumes that are too small to exhibit spontaneous symmetry breaking.
The finite-volume symmetry restoration occurs for smaller and smaller values of the coupling as the volume is increased, which is consistent with it being a finite-volume effect.

Fig.~\ref{fig:OS observable data} also shows a fit of the data from the largest volume to a functional form $f(e^2)= A \exp{(-B/e^2)}$ over the range of couplings $0.50 \leq e^2 \leq 0.80$; these couplings are large enough to avoid symmetry restoration from the finite volume and are small enough to remain close to the continuum limit. The fit form is inspired by the analytic results for the standard $\U(1)$ theory (for example Eq.~\eqref{eq:mass gap}) as well as by the effective theory for the $\extParam=\pi$ theory (see Eq.~\eqref{eq:mass staggered prediction}) and provides an excellent fit to the data. Such a functional form would also imply that the $S$ symmetry remains broken down to the continuum $e^2 \to 0$. In order to test whether the symmetry is restored at some small, but non-zero value of $e^2$, we also introduced an offset in the fit form, which was therefore modified to $f(e^2)= A \exp{(-B/(e^2-e^2_c))}$, giving a value $e^2_c = 0.04(11)$, which is consistent with zero.
Overall, we interpret the numerical evidence to mean that the $S$ symmetry is broken for a wide range of couplings, possibly down to $e^2 \to 0$ although we cannot strictly exclude that it could be restored at some small but non-zero $e^2$. Importantly, we note that even though the single-site shift symmetry $S$ is broken, translation symmetry by an even number of lattice spacings remains unbroken, and we therefore expect to recover full translational invariance in the continuum.

\begin{figure*}
    \centering
    \includegraphics{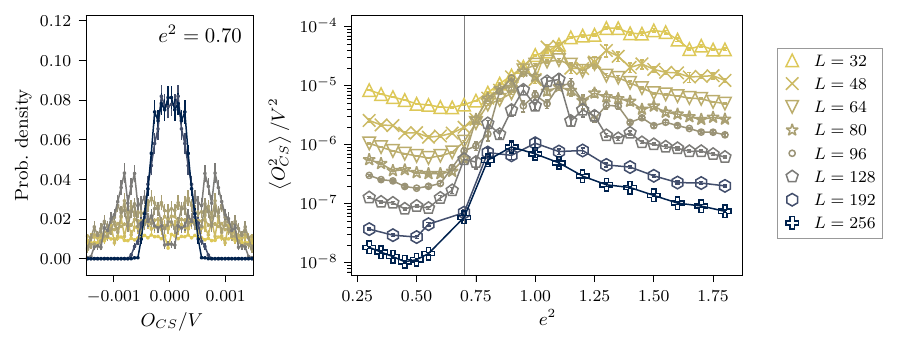}
    \caption{Left: Histogram of the operator $O_{CS}$ normalized by the volume at $e^2=0.70$. Right: Susceptibility of $O_{CS}$ normalized by $V^2$ as a function of $e^2$ for several volumes. The vertical line is located at the coupling $e^2=0.70$ where the histogram is shown. A single histogram peak at large volumes (left) and scaling smaller than $V^2$ (right) indicates that either the $C$ or $S$ symmetries are unbroken.}
    \label{fig:OCS observable data}
\end{figure*}

The situation is quite different for the observable $O_{CS}$, whose numerical data is shown in Fig.~\ref{fig:OCS observable data}. In this case the normalized susceptibility $O_{CS}^2/V^2$ decreases with the volume for all couplings considered, indicating that $O_{CS}$ does not acquire a vacuum expectation value. The decrease is less clear in the central region, $0.70 \lesssim e^2 \lesssim 0.80$, especially for the larger volumes. However, the histogram for $e^2 = 0.70$ shows that while two peaks appear to be forming for small volumes, they merge into a single peak around zero as the volume is increased. Therefore the observable $O_{CS}$ does not acquire a vacuum expectation value at any value of the coupling, which means that at least one of $C$ or $S$ remains unbroken. Since we have seen that $S$ is broken, the data implies that charge conjugation $C$ remains unbroken for all values of the couplings studied.

\subsection{Mass}\label{sec:mass}
The mass and string tension for the staggered height model were also measured. These are particularly interesting because they can be compared with the effective theory prediction which we derive in Section \ref{sec:effective theory} as well as with the analytical and numerical results for the standard $\U(1)$ theory. 

The mass may be measured from the exponential decay of
correlations between height variables $h_x$ and $h_y$ with increasing distance,
after appropriate momentum projection and subtractions \cite{FrRandomSurf,Evertz}. The correlation function $(h_x-h_y)^2$ is appropriately invariant under the global $\Z$-symmetry. However, since it does not factorize into a product of operators which are local in time, one may worry that it does not admit a proper spectral interpretation. In fact, while each of the three terms $(h_x-h_y)^2 = h_x^2 + h_y^2 - 2h_x h_y$ admits a proper spectral interpretation, their individual expectation values are not well-defined because they are not $\Z$-invariant. Introducing transfer-matrix eigenstates $\{ \ket{n} \}$, on a lattice of time extent $T$ one finds for $T \to \infty$, 
\begin{equation}
\begin{aligned}
    &\expval{(h_x-h_y)^2} = \\
    &\quad \bra{0} h_x^2 \ket{0} + \bra{0} h_y^2 \ket{0} -2 \bra{0} h_x \ket{0} \bra{0} h_y \ket{0}\\
    &\quad -2 \sum_{n > 0} \bra{0} h_x \ket{n} \bra{n} h_y \ket{0} e^{-(E_n-E_0)t} \ ,
\end{aligned}
\end{equation}
where $t$ is the time separation between $x$ and $y$.  The  expectation values in the first line are not individually $\Z$-invariant, but the $\Z$-dependence cancels between these terms, leaving a simple $t$-independent vacuum contribution.
On the other hand, the matrix elements $\bra{0} h_x \ket{n}$ are $\Z$-invariant for $n \neq 0$ and are therefore well-defined. By subtracting the constant vacuum contribution, the exponential decay of the correlation function $(h_x-h_y)^2$ thus probes the non-vacuum states in the symmetry sector of the operator $h_x$.

The mass is extracted from the same numerical simulations based on the cluster algorithm as used for the order parameters above. It is important to note that we are primarily interested in the behavior of this mass in the phase with the correct symmetry structure. In particular, as shown in Fig.~\ref{fig:OS observable data}, the broken $\Z_2$ symmetry is restored at small couplings due to finite volume effects. Working at a certain fixed volume $L^3$, we therefore limit the range of couplings included in fits to those where the symmetry remains broken, though the mass has been measured for all simulated couplings.

\begin{figure}
    \centering
        \includegraphics[width=\linewidth]{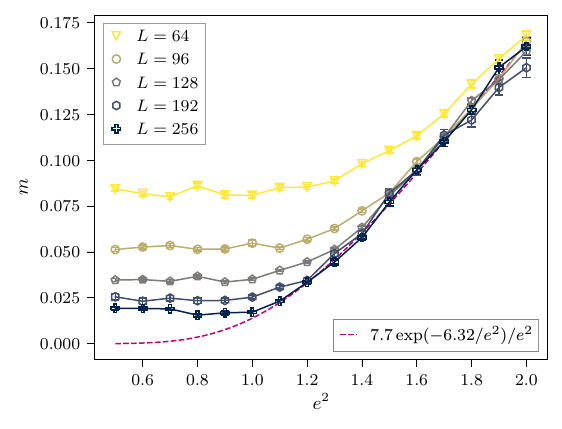}
    \caption{Mass $m$ of the $\extParam=\pi$ theory for various values of the coupling $e^2$, given in lattice units. Finite volume effects can be seen from the measurements at smaller volumes diverging from larger volumes as $e^2$ is taken small. The fit is performed with data in the range $e^2 \in [1.1, 2.0]$, with the fit form motivated by our effective theory for $\extParam=\pi$ (Eq.~\eqref{eq:mass staggered prediction}). }
    \label{fig:mass numerical results}
\end{figure}

The results of the numerical simulations for the mass of the $\extParam=\pi$ theory are shown in Fig.~\ref{fig:mass numerical results} for $L=256$ and a range of couplings, together with a fit to an exponential form. Simulations were performed for several volumes; this allows us to establish that the points in the range $e^2 \in [0.6, 1.0]$ suffer from finite-volume effects and are therefore excluded from the fit. The mass of the $\extParam=\pi$ theory decreases much more quickly as $e^2$ is decreased compared to the standard theory. In fact, according to the effective theory for $\extParam=\pi$ (Eq.~\eqref{eq:mass staggered prediction}), the constant in the exponent should be twice as large as compared to the standard theory. We have attempted to fit our data for the largest volume in the range $[1.1, 2.0]$ where it doesn't suffer from finite-volume effects. For the fit, we chose a function of the form
\begin{equation}
    f(e^2) = A g(e^2) \exp{\pqty{-B/e^2}} \ ,
\end{equation}
where $g$ represents different choices of prefactor. We considered $g(e^2)=1$ (simple exponential decay), $g(e^2)=1/e^2$ (effective theory prediction, Eq.~\eqref{eq:mass staggered prediction}), and $g(e^2) = 1/\sqrt{e^2}$ (behaviour of the standard theory, Eq.~\eqref{eq:mass gap}). All three choices provide acceptable fits to the data, with the effective theory prediction slightly favored. The fits have a $\chi^2/\mathrm{d.o.f}\approx 0.80, 0.95, 0.78$, and exponent $B = 4.85(7), 6.33(7), 5.59(7)$ respectively. As such, our data is in qualitative agreement with the effective theory prediction of $2\pi^2 v_0 \approx 4.99$ in Eq.~\eqref{eq:mass staggered prediction} for the exponent $B$. However, the data is not sufficiently precise to distinguish the three cases for $g(e^2)$, and we are working at a range of couplings which is sufficiently far from the continuum where corrections to scaling are likely important. 
Similarly to what was done for the observable $O_S$ in Section \ref{sec:order parameters}, we have also attempted to introduce an offset by replacing $e^2$ with $e^2 - e_c^2$ in the fit forms, in order to test the hypothesis that the transition takes place at a non-zero critical value of the coupling $e_c^2 \neq 0$. In all cases, we have found the offset $e_c^2$ to be consistent with zero within error.

\subsection{Energy-momentum dispersion relation}
We next consider analogous correlation functions at non-zero momentum. The operator
\begin{equation}
    h(\vec{k},t) \equiv \sum_{\vec{x}} e^{i \vec{x} \cdot \vec{k}} h(\vec{x},t)
\end{equation}
creates a state at non-zero momentum $\vec{k}$ as long as it is compatible with the finite-volume quantization condition $\vec{k} = \frac{2\pi}{L}(n_1, n_2)$ with $n_{1,2} \in \mathbb{Z}$. For all $\vec{k} \neq \vec{0}$, the operator $h(\vec{k},t)$ is $\Z$-invariant. In this case, we are free to analyze the relatively simpler correlation functions
\begin{equation}
    C(\vec{k},t) \equiv \braket{h(-\vec{k},t) h(\vec{k},0)} \ ,
\end{equation}
which require no vacuum subtraction. Fitting these correlation functions to single exponentials yields an estimate of the energy $E(\vec{k})$ of the lowest lying state with given momentum $\vec{k}$.

\begin{figure}
    \centering
    \includegraphics{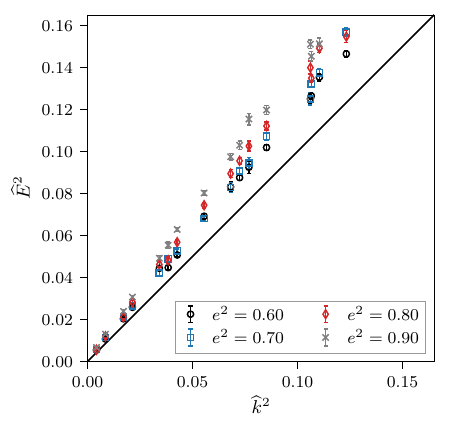}
    \caption{Measurements of the energy-momentum dispersion relation for a range of couplings, plotted using the lattice-units quantities $\widehat{E}^2 = 4 \sinh(a E / 2)^2$ and $\widehat{k}^2 = 4 - 2 \sum_i \cos(a k_i)$ inspired by the dispersion relation of a free relativistic particle on the lattice. In the continuum limit of a relativistic theory, the dispersion relation is expected to converge to the limit $\widehat{E}^2 = \widehat{k}^2$ indicated by the black line. A clear trend towards this relativistic relation can be observed as the bare coupling is decreased towards the continuum limit. All measurements were performed on an $L = 96$ lattice volume.}
    \label{fig:E_vs_k}
\end{figure}

To study $E(\vec{k})$, we performed measurements across a variety of momenta for four choices of the bare coupling $e^2$ at $L=96$, making sure that our choices of couplings all lie within the correct symmetry phase for the volume used (see Fig.~\eqref{fig:OS observable data}). Fig.~\ref{fig:E_vs_k} plots the lattice-units quantities $\widehat{E}^2 = 4 \sinh(a E / 2)^2$ and $\widehat{k}^2 = 4 - 2 \sum_i \cos(a k_i)$ which give lattice approximations to $a^2E^2$ and $a^2\vec{k}^2$ inspired by the simple lattice discretization of a relativistic free particle. In the continuum limit, these are expected to satisfy $\widehat{E}^2 = \widehat{k}^2$. This reference relation is shown by the diagonal line in the plot, and the data can be seen to approach this scaling as the coupling $e^2$ is taken smaller. Further measurements would be required to reliably extrapolate to the continuum relation, but the data shown already strongly suggest the emergence of the full $O(3)$ spacetime symmetry group corresponding to a Euclidean representation of a relativistic theory in the continuum limit.

\subsection{String tension} \label{sec:string tension}

To measure the string tension, we choose to insert a pair of static charges, one positive and one negative, directly in the partition function. We then employ two complementary methods: one based on a direct measurement of the system's energy with the charges inserted \cite{QuantumLink, sterling-greensite}
and another based on the snake algorithm \cite{deForcrand:2000fi, deForcrand:2004jt}.

\begin{figure}
    \centering
    \includegraphics[]{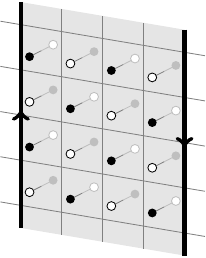}
    \caption{An example of the relation between a Polyakov loop (bold lines), the corresponding surface in the original lattice (gray surface), and the dual links $l$ (links between black/white dual sites) for which $s_l$ is non-zero. The orientation of the Polyakov loops gives an orientation to the surface, determining the signs for the dislocation variables $s_l$. For clarity, only a two-dimensional slice of the 3D volume is shown.}
    \label{fig:polyakov_corr}
\end{figure}

In both cases, a static charge propagating in time at spatial position $\vec x$ is represented in the original theory by a Polyakov loop $P(\vec x)$ wrapping around the time direction, i.e.
\begin{equation}
    P(\vec x) = \prod_{t=0}^{T-1} e^{i \varphi_t(t,\vec{x})} \ ,
\end{equation}
where $\exp{\pqty{i \varphi_t(t,\vec{x})}}$ is the $\U(1)$ link variable at position $(t,\vec{x})$ oriented in the time direction. The expectation value of the correlator of a charge pair may be expressed directly in the dual theory by \cite{GopfMack}
\begin{multline}
    \expval{P(\vec x_1)^* P(\vec x_2)} = \frac{1}{Z} \pqty{\prod_{x \in \mathrm{sites}} \sum_{h_x}} \\ \times \exp{\bqty{-\frac{e^2}{2} \sum_{\expval{xy}} (h_x-h_y+s_{\expval{xy}})^2}} \ ,
\end{multline}
where all quantities on the right-hand side live on the dual lattice. Here $s_l$ is a field of `dislocations' defined on links of the dual lattice by
\begin{equation} \label{eq:sl}
    s_l = \begin{cases} \pm 1 & \prescript{\star}{}{l} \in A\\ 0 & \mathrm{otherwise} \end{cases} \ ,
\end{equation}
where $\prescript{\star}{}{l}$ is the plaquette in the original lattice dual to link $l$, and $A$ is any surface (i.e.\ connected collection of plaquettes in the original lattice) bounded by the two Polyakov loops at spatial positions $\vec x_1$ and $\vec x_2$. The sign of $s_l$ is positive or negative depending on the orientation in which the link $l$ is traversed. The expectation value $\expval{P(\vec x_1)^* P(\vec x_2)}$ is independent of deformations of the surface $A$, and for convenience in simulations we choose $A$ to be the rectangular surface bounded by the Polyakov loops at $\vec{x}_1$ and $\vec{x}_2$ which does not cross the boundary. An example of the geometry of the Polyakov loops, rectangular surface $A$, and the corresponding dual links is shown in Fig.~\ref{fig:polyakov_corr}.

There also exist topologically inequivalent surfaces bounded by the same Polyakov loops: for example, the surface starting at $\vec{x}_1$ and wrapping backwards through the lattice boundary to $\vec{x}_2$ cannot be obtained by local deformations of our choice of surface which goes forwards from $\vec{x}_1$ to $\vec{x}_2$ without crossing the boundary. The choice between these topologically inequivalent surfaces amounts to a choice of boundary conditions in the dual theory, which are equivalent up to finite-volume effects. Our choice of surface allows us to suppress the wrap-around contributions and get more precise estimates of the string tension using separations $|\vec{x}_1 - \vec{x}_2| \geq L/2$ in the following.

The Polyakov loop correlator $\expval{P(\vec x_1)^* P(\vec x_2)}$ can be interpreted as the ratio of two partition functions, one with the static charge pair and one in the vacuum sector. To do this, we define the generalized partition function $Z[s]$ in the background of the dislocations $s_l$ by
\begin{equation}
\begin{aligned}
    Z[s] &= \pqty{\prod_{x \in \mathrm{sites}} \sum_{h_x}} \\
    &\times \exp{\bqty{-\frac{e^2}{2} \sum_{\expval{xy}} (h_x-h_y+s_{\expval{xy}})^2}} \ .
\end{aligned}
\end{equation}
Note that $Z[0] = Z$ coincides with the partition function of the staggered height model Eq.~\eqref{eq:staggered height model}, and we then have
\begin{equation}
\expval{P(\vec x_1)^* P(\vec x_2)} = Z[s] / Z[0] \ ,
\end{equation}
for $s$ defined as in Eq.~\eqref{eq:sl}.

Variations on the `snake algorithm' have been introduced to effectively evaluate such ratios of partition functions~\cite{deForcrand:2000fi,deForcrand:2004jt}. The basis of these approaches is to measure ratios of partition functions $Z[s_{n+1}] / Z[s_n]$ using independent Monte Carlo calculations, finally giving the desired ratio $Z[s] / Z[0]$ by a telescoping product. We adopt an analogous simulation strategy, using a series of Monte Carlo simulations to evaluate the ratios of Polyakov loop correlators separated by a variety of spatial distances aligned along the $x$-axis (the $\hat{1}$ direction) of the lattice. Each Monte Carlo evaluation is constructed to evaluate
\begin{equation}
    \frac{\braket{P((R+1) \, \hat{1} + \vec{x}_0)^* P(\vec{x}_0)}}
    {\braket{P(R \, \hat{1} + \vec{x}_0)^* P(\vec{x}_0)}} \equiv \frac{Z[s_{R+1}]}{Z[s_R]} \ ,
\end{equation}
for some spatial separation $R$. The fields $s_{R+1}$ and $s_R$ are determined by the Polyakov loop geometries, as discussed above, and only differ in whether they include the column of dual links corresponding to the plaquettes between spatial sites $R \, \hat{1} + \vec{x}_0$ and $(R+1) \, \hat{1} + \vec{x}_0$.

To gain further insight into the nature of the confining string, we also adopt a second simulation strategy to measure the energy of the static charges based on the Hamiltonian of the system.
Since the partition function $Z[s]$ may be expressed in terms of a Hamiltonian $H$ via the relation $Z[s] = \tr(e^{-\beta H})$, it is possible to obtain $\expval{H}$ by analytically differentiating $Z[s]$ with respect to $\beta$. One then obtains $\expval{H}$ as an observable that is amenable to Monte Carlo simulation with respect to the probability distribution defined by $Z[s]$, in particular
\begin{multline}
    \label{eq:energy operator}
    \expval{H} = \frac{e^2}{2 T} \bigg\langle- \sum_{\expval{xy}_{\mathrm{time}}}(h_x-h_y+s_{\expval{xy}})^2+\\
    +\sum_{\expval{xy}_{\mathrm{space}}} (h_x-h_y+s_{\expval{xy}})^2\bigg\rangle \ ,
\end{multline}
where $\expval{xy}_{\mathrm{time}}$ and $\expval{xy}_{\mathrm{space}}$ denote links in the time and space directions respectively. A careful derivation of this result is presented in Appendix \ref{sec:charge insertion energy derivation}. 
One expects Eq.~\eqref{eq:energy operator} to be valid only in the continuum limit, but it can still be used to give qualitative information about the structure of the confining string and to cross-check the string tension close to the continuum limit. In particular, an advantage of this formulation is that one may also measure the local energy contribution by any individual link and thus visualize the energy distribution of the confining string. As a simulation strategy, the energy \eqref{eq:energy operator} is measured using Monte Carlo evaluations at several particle-antiparticle separations $R$, obtaining $E(R)$ as a function of $R$. These Monte Carlo estimates are given by separate Metropolis simulations with fixed static charge pairs for each separation.

From either approach, the string tension $\sigma$ has then been extracted by fitting
\begin{equation}
    E(R) = -\frac{1}{T} \ln\braket{P^*(\vec{x}_1) P(\vec{x}_2)}\big|_{|\vec{x}_1 - \vec{x}_2| = R}
\end{equation}
to the theoretical predictions of the energy of the confining string, which for large separations $R \gg 1/\sqrt{\sigma}$ is given by \cite{LSW1980,arvis,LuscherWeisz,luscher1981,stack-stone}
\begin{equation}
    E(R) = A + \sigma R - \frac{B}{R} \ + \mathcal{O}(1/R^2) \ .
\end{equation}
Alternatively, the estimates of the string tension can be directly obtained from the snake algorithm approach, which most directly gives access to finite differences of the energy
\begin{equation}
    E(R+1) - E(R) = \sigma + \mathcal{O}(1/R^2) \ .
\end{equation}
We have also used this method to obtain precise estimates of $\sigma$ from the snake algorithm without fitting and without needing to perform Monte Carlo evaluations for all $R$. The two methods give consistent results when compared on identical lattice parameters, and we therefore adopt the latter approach for all following measurements of $\sigma$, averaging estimates from several choices of $R \approx L/2$ to minimize the higher-order effects.

\begin{figure}
    \centering
    \includegraphics{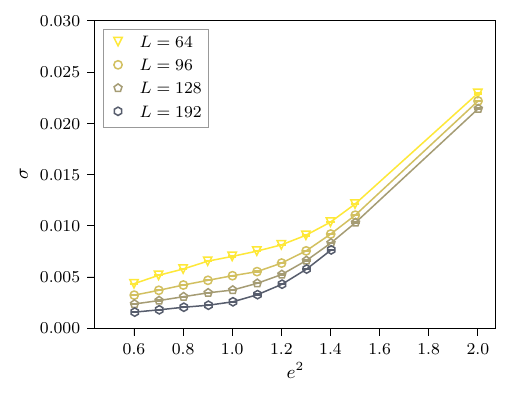}
    \caption{String tension $\sigma$ of the $\extParam=\pi$ theory for various values of the coupling $e^2$, given in lattice units. Statistically significant finite-volume effects can be observed over the entire range of couplings, potentially reflecting the more diffuse structure of the double-stranded confining string in this theory.}
    \label{fig:string tension numerical results}
\end{figure}

In order to test this method, we have used it to compute the string tension in the standard three-dimensional $\U(1)$ gauge theory, and we have obtained results compatible with both the analytical predictions Eq.~\eqref{eq:string tension analytical} and results previously published in the literature \cite{TepAthen}. The static charge potentials measured by the snake algorithm and the Hamiltonian method were also compared and found to agree with improving accuracy as $e^2$ was taken small, corresponding to the continuum limit where both definitions are the same.

\begin{figure*}
    \centering
    \begin{tabular}{ll}
    \includegraphics[height=6.5cm]{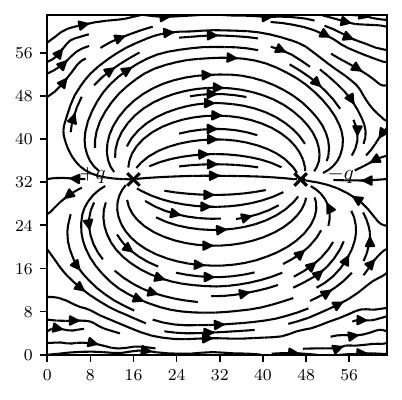} &
    \includegraphics[height=6.5cm]{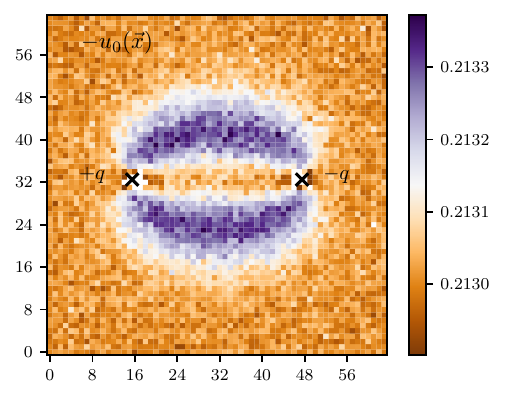} \\
    \includegraphics[height=6.5cm]{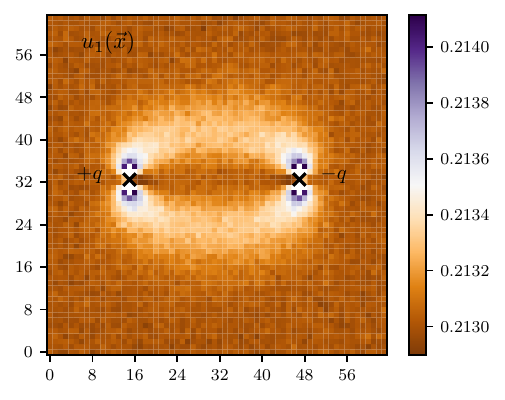} &
    \includegraphics[height=6.5cm]{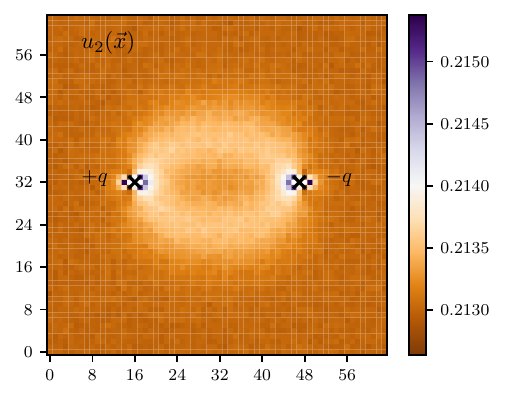}
    \end{tabular}
    
    \caption{Time-averaged local energies and electric field on a spatial $L \times L$ lattice measured in the presence of two static charges on the ensemble with coupling $e^2=2.0$ and size $L=64$. The first figure (top left) shows the field lines of the electric field $\vec{E}$ defined in Eq.~\eqref{eq:electric field}. The last three figures (top right through bottom right) show the local energy defined in Eq.~\eqref{eq:energy density} in the three spacetime directions. In all three figures, the brightest pixels corresponding to contact energy terms are masked. The energy distributions clearly depict fractionalization of the flux string into two strands.}
    \label{fig:local string structure}
\end{figure*}

The results of the numerical simulations for the string tension are shown for the $\extParam = \pi$ theory in Fig.~\ref{fig:string tension numerical results} for $L \in \{64, 96, 128, 192\}$. The string tension clearly decreases in lattice units towards the continuum limit, as expected. In contrast to the mass, however, one sees statistically significant finite volume effects throughout the range of couplings studied. We have also compared to simulations of the ordinary $U(1)$ gauge theory with $\extParam = 0$, in which case the finite volume effects are less significant.

The structure of the string in the $\extParam = \pi$ theory provides a potential explanation for this feature. To investigate this structure, we consider the local electric field
\begin{equation} \label{eq:electric field}
    \vec{E}(x) = ( h_x - h_{x+\hat{2}} + s_2(x) , - h_x + h_{x+\hat{1}} - s_1(x)) \ ,
\end{equation}
where the dislocation $s_\mu(x) = s_l$ on the link $l$ connecting $x$ and $x + \hat{\mu}$, and local energy density terms
\begin{equation} \label{eq:energy density}
\begin{aligned}
    u_0(x) &= \expval{ - (h_x-h_{x+\hat{0}} + s_0{(x)})^2}  \ , \\
    u_i(x) &= \expval{ (h_x-h_{x+\hat{i}} + s_i{(x)})^2} \ ,
\end{aligned}
\end{equation}
which together define the Hamiltonian energy density $H(x) = u_0(x) + u_1(x) + u_2(x)$ in Eq.~\eqref{eq:energy operator}.
Fig.~\ref{fig:local string structure} depicts an example of the measured energy density terms and electric field for the coupling $e^2 = 2.0$ and a particular separation of the static charge and anti-charge.
The electric field lines have the typical structure associated with a positive and negative charge pair. On the other hand, the energy density depicts a clear fractionalization of the confining string into two strands for the $\extParam = \pi$ theory. We have observed a similar fractionalized structure for all couplings considered, however the choice of coupling $e^2=2.0$ results in the clearest pictures. This string fractionalization suggests that the linear scaling regime of the static charge potential may only set in at larger physical distances, explaining the relatively strong finite-volume effects observed in the string tension.

\begin{figure*}
    \centering
    \includegraphics[height=6.5cm]{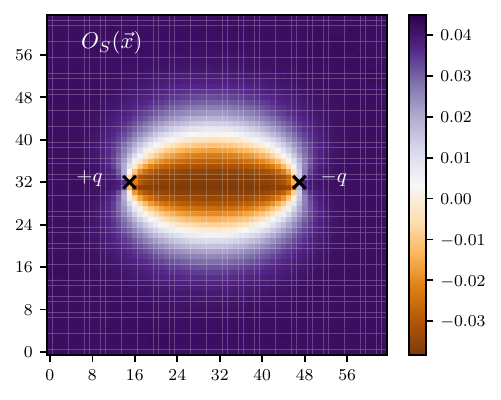}
    \quad
    \includegraphics[height=6.5cm]{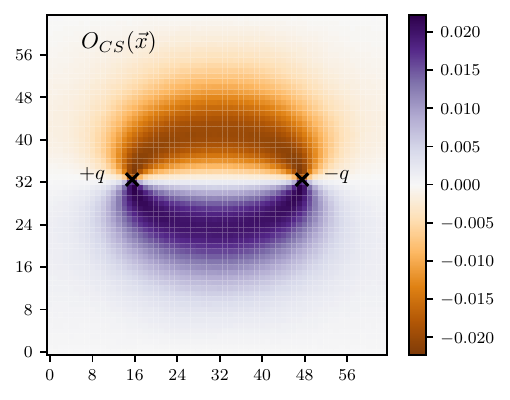}

    \caption{Local $O_{S}$ (left) and $O_{CS}$ (right) observables measured in the presence of two static charges on the ensemble with coupling $e^2=2.0$ and lattice size $L=64$. Shown are the Euclidean-time-averaged measurements over the spatial $L \times L$ lattice.
    The $O_S$ observable shows that the two string strands separate regions that correspond to the two ground states of the broken $\Z_2$ symmetry, while the $O_{CS}$ observable shows that the two strands themselves break charge conjugation.}
    \label{fig:local observables}
\end{figure*}

The confining string also shows a non-trivial interplay with the phase structure of the theory. We define local versions of the order parameters $O_{CS}$ and $O_S$, which are invariant under the global $\Z$-symmetry and under deformations of the arbitrary surface connecting the two Polyakov loops. For $O_{CS}$ we define
\begin{equation}
    O_{CS}(x) = \frac{(-1)^x}{12} \sum_{\hat{\mu} \in \{\pm \hat{0},\pm \hat{1}, \pm \hat{2} \} } (h_x-h_{x + \hat{\mu}} + s_{\mu}(x)) \ .
\end{equation}
While the expression is complicated, one can also define a local version $O_S(x)$ of the operator $O_S$. In order to make it invariant under deformations of the surface bounded by the Polyakov loops, one rewrites it in terms of link variables $h_x - h_{x+\mu}$ and then replaces $h_x - h_{x+\mu} \to h_x - h_{x+\mu}+s_{\mu}(x)$. In the case where there are no static charges, summing over these local observables recovers the original order parameters, 
\begin{equation}
    O_{CS} = \sum_x O_{CS}(x) \ , \qquad O_{S} = \sum_x O_{S}(x) \ .
\end{equation}
To understand the relation between the order parameter structure and the broken symmetry, we perform measurements using an ensemble where no tunneling events occurred, which would otherwise wash out the figure.

Fig.~\ref{fig:local observables} shows the resulting distributions of the observables $O_{S}(x)$ and $O_{CS}(x)$ in the presence of the flux string. The $O_S(x)$ observable shows that inside and outside the two strands the system is found in two different ground states of the broken $\Z_2$ symmetry $S$. This indicates that the individual strands of the fractionalized string play the role of domain walls between the two vacua of the spontaneously broken symmetry.
Meanwhile, the $O_{CS}(x)$ observable takes non-zero values within each strand of the fractionalized string, which is consistent with explicit breaking of $C$ symmetry by the insertion of static charges with particular positive and negative values. This order parameter remains zero outside the static charge system as expected from the lack of symmetry breaking observed in the vacuum sector for $O_{CS}$. In both cases, the particular signs of the order parameters where they are non-zero is not fixed, but spontaneously selected; they are flipped by tunneling events between the vacua of the spontaneously broken $S$ symmetry.

\subsection{Mass and string tension scaling}
A key prediction for both the ordinary $\extParam=0$ theory (see Eq.~\eqref{eq:string tension analytical}) and the $\extParam=\pi$ theory (see Section \ref{sec:effective theory}) is that (in units where $a = 1$) $\sigma$ should scale as $m e^2$. This implies the three possible continuum limits discussed in Sec.~\ref{sec:usual u(1)}. By comparing our mass and string tension data, we could attempt to validate this scaling prediction. However, the strong finite volume effects in both the mass and string tension make it difficult to reliably determine an infinite volume ratio $\sigma / m e^2$ to be extrapolated towards the continuum limit.

Instead, we performed an additional set of Monte Carlo calculations extrapolating along a line of constant physics in a fixed physical volume, as defined by fixing $mL$ to a constant value. As long as $mL \gg 1$, we expect the argument for the scaling of string tension with mass to still hold. Thus by studying these constant physical volume ensembles, we can determine whether such ratio has a finite continuum  limit. To perform this study, we used five different lattice sizes with couplings tuned to fix $mL$ to $6$, $8$, and $10$, as detailed in Table~\ref{tab:mL-tuning}. We use a prescription of extracting the string tension by the snake algorithm approach described above with measurements of the energy differences based on a window of separations $R \approx 5 L / 8$. This value was chosen to maximally suppress the $O(1/R^2)$ terms while avoiding the boundary of the lattice. Finally, we use the fact that (in lattice units) the ratio can be written as
\begin{equation}
    \frac{\sigma}{m e^2} = \frac{\sigma L}{e^2 (mL)}.
\end{equation}
Using the fixed choice of $mL$ instead of direct measurements of the mass allows a much more precise determination of this ratio, although this neglects systematic uncertainties from mistunings of the mass.

\begin{table}
    \centering
    \begin{ruledtabular}
    \begin{tabular}{cc ccccc c}
        \hspace{.25cm} \multirow{2}{*}{$mL = 6$}\hspace{.25cm} & $L$ \hspace{.5cm} & $64$ & $96$ & $128$ & $192$ & $256$ & \hspace{.25cm} \\
        &$e^2$ \hspace{.5cm} & $1.35$ & $1.29$ & $1.23$ & $1.12$ & $1.09$ & \\[1ex]
        \hline
        \hspace{.25cm} \multirow{2}{*}{$mL = 8$}\hspace{.25cm} & \rule{0pt}{3.5ex}$L$ \hspace{.5cm} & $64$ & $96$ & $128$ & $192$ & $256$ & \hspace{.25cm} \\
        &$e^2$ \hspace{.5cm} & $1.68$ & $1.50$ & $1.39$ & $1.24$ & $1.18$ & \\[1ex]
        \hline
        \hspace{.25cm} \multirow{2}{*}{$mL = 10$}\hspace{.25cm} & \rule{0pt}{3.5ex}$L$ \hspace{.5cm} & $64$ & $96$ & $128$ & $192$ & $256$ & \hspace{.25cm} \\
        &$e^2$ \hspace{.5cm} & $1.92$ & $1.65$ & $1.49$ & $1.33$ & $1.25$ &
    \end{tabular}
    \end{ruledtabular}
    \caption{Choices of coupling constant $e^2$ and lattice size $L$ tuned over a variety of lattice spacings to fix $mL$ to $6$, $8$, and $10$.}
    \label{tab:mL-tuning}
\end{table}

\begin{figure}
    \centering
    \includegraphics{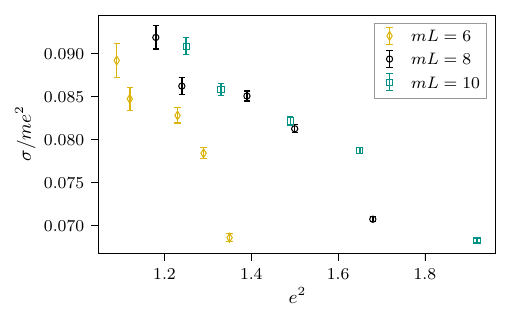}
    \caption{Measured ratios $\sigma / me^2$ across a range of ensembles tuned to fix $mL$ to $6$, $8$, and $10$, given in lattice units. The resulting ratios can be seen to plateau towards small $e^2$, corresponding to smaller lattice spacings, but a notable rise is still present at the smallest lattice spacings accessible. Nevertheless, the measured ratios are quite distinct from estimates of the scaling ratio $\widetilde{c}/4\pi^2 \approx 0.21$ in the $\extParam = 0$ theory~\cite{TepAthen}.
    }
    \label{fig:sigma-m-ratios}
\end{figure}

The resulting measurements of the ratios are shown in Fig.~\ref{fig:sigma-m-ratios}. At the available lattice spacings, the ratio determined for each of the fixed physical volumes can be seen to approximately plateau. Nevertheless, at the smallest $e^2$, corresponding to the smallest lattice spacing, the data continue to rise indicating that even finer lattices would be required to confirm the scaling behavior.

\subsection{Discussion}

Our results provide evidence for a continuum limit of the $\extParam=\pi$ theory which is distinct from that of the $\extParam = 0$ theory, despite both theories having exact $U(1)$ gauge symmetry, charge conjugation symmetry, macroscopic translational symmetry, and cubic symmetry. In particular, the order parameters suggest that the $\extParam=\pi$ theory has a continuum limit obtained as $e^2 \to 0$ where the $\Z_2$ symmetry $S$ is spontaneously broken and charge conjugation symmetry is unbroken. This theory also shows qualitatively different behaviour, most strikingly in the fractionalization of the confining flux string. This behaviour is confirmed by the effective theory for the $\extParam=\pi$ theory which we derive in Section \ref{sec:effective theory}. The data for the mass and the string tension are also compatible with the effective theory predictions, although better control of the finite volume effects and the corrections to scaling is required in order to test quantitative agreement.

Both the string fractionalization and the phase separation have been previously observed in the context of quantum link models \cite{QuantumLink,QuantumDimer1,QuantumDimer2,NematicTriangular}. In the results observed for some such quantum link models,
a spontaneous breaking of parity symmetry was also observed in the static charge system, so that the strands of the string formed an asymmetric arrangement. On the other hand, our results for the $\extParam=\pi$ theory do not demonstrate any such parity breaking.

\section{Effective theory}\label{sec:effective theory}

In order to better understand the behavior of the $\extParam=\pi$ theory and confirm our numerical results, in this section we derive an effective theory following \cite{GopfMack}. As we have seen in Section \ref{sec:usual u(1)}, the standard Abelian gauge theory in three dimensions has been shown to be equivalent to a Sine-Gordon model with action
\cite{GopfMack, SemiAbelian, Banks:1977cc, Muller1981}
\begin{multline}
    \label{eq:usual u(1) effective theory 2}
    S[\phi] = \frac{1}{2} \sum_{\expval{xy}} (\phi_x-\phi_y)^2 - 2 \lambda\sum_x \cos{\pqty{\frac{2\pi\phi_x}{\sqrt{e^2}}}} \ ,
\end{multline}
where $\phi_x$ is a real scalar field and $\lambda \equiv 2e^{-2\pi^2 v_0 /e^2}$. This effective theory is valid for small $e^2$, where the continuum limit is expected to lie. The derivation leading to Eq.~\eqref{eq:usual u(1) effective theory 2} can be performed also in the case of our staggered height model, and the corresponding effective theory is given again by a Sine-Gordon model with a \textit{staggered} potential,
\begin{equation}
    \label{eq:staggered cosine model}
    S[\phi] = \frac{1}{2} \sum_{\expval{xy}} (\phi_x-\phi_y)^2 - 2 \lambda\sum_x (-1)^x \cos{\pqty{\frac{2\pi\phi_x}{\sqrt{e^2}}}} \ .
\end{equation}
Non-relativistic versions of Eq.~\eqref{eq:staggered cosine model} have been discovered as effective descriptions of quantum antiferromagnets \cite{ReadSachdev}. Due to the staggering, this model, unlike the action in Eq.~\eqref{eq:usual u(1) effective theory 2}  does not immediately admit an analytical continuum limit. We therefore rewrite it in terms of variables which, as we will see, allow a continuum description. Following the discussion in \cite{ReadSachdev}, we define sum and difference variables 
\begin{align}
    \label{eq:sum and difference variables}
    \chi_x = \frac{\phi_x + \phi_{x+\hat\mu}}{2} \ , \qquad
    \xi_x = \frac{\phi_x - \phi_{x+\hat\mu}}{2}  \ ,  
\end{align}
defined to live only on the \textit{even} lattice sites. In particular, this definition singles out a specific direction $\mu$ and combines the scalar fields on the even site $x$ and odd site $x+\hat\mu$ into the sum and difference variables on the even sites. This mapping is one-to-one and has a unit Jacobian. The arbitrary choice of direction disappears in the final result. Substituting into the effective theory, we find that $\xi$ has a mass of the order of the lattice cutoff: we therefore ignore all of its gradients \cite{ReadSachdev} and find 
\begin{multline}
    \label{eq:staggered cosine model rewriting}
    S[\chi,\xi] = \sum_x \bigg[ \frac12 (\vec \nabla \chi_x)^2 + 12 \xi_x^2 + \\+ 4 \lambda \sin{\pqty{\frac{2\pi\chi_x}{\sqrt{e^2}}}} \sin{\pqty{\frac{2\pi\xi_x}{\sqrt{e^2}}}} \bigg]  \ .
\end{multline}
In Eq.~\eqref{eq:staggered cosine model rewriting} the sum is taken only over the even sites of the original hypercubic lattice. The corresponding lattice, the one on which the effective theory in Eq.~\eqref{eq:staggered cosine model rewriting} is defined, is therefore a three-dimensional lattice of tetrahedra and octahedra; each point is connected to nearest neighbours in four directions, and in particular $(\vec \nabla \chi_x)^2$ is the sum of the squares of the lattice derivatives in the four directions. Since it is an isotropic kinetic term on the lattice, we expect that it will become a Lorentz-invariant kinetic term in the continuum limit. Next, we again use the fact that
$\xi$ has a mass of the order of the lattice cutoff to integrate it out. Its mean-field value is given by \cite{ReadSachdev}
\begin{equation}
    \xi_x \approx -\frac{\pi}{3} \frac{\lambda}{\sqrt{e^2}} \sin{\pqty{\frac{2\pi\chi_x}{\sqrt{e^2}}}} \ ,
\end{equation}
which we substitute back into the action to find
\begin{equation}
    \label{eq:final effective theory}
    S[\chi] = \frac{e^2}{4\pi^2} \sum_x \bqty{ \frac12 (\vec \nabla \chi_x)^2 + \frac{8}{3} \pi^4 \frac{\lambda^2}{e^4} \cos{\pqty{2\chi_x }} } \ ,
\end{equation}
where we also redefined $\chi_x \to \frac{\sqrt{e^2}}{2\pi} \chi_x$ for clarity.

As is clear from Eqs.~\eqref{eq:staggered cosine model} and \eqref{eq:staggered cosine model rewriting}, this model admits a global $\Z$ symmetry which acts as $\chi_x \to \chi_x + 2\pi$. We identify this symmetry with the global $\Z$ redundancy of the original model, meaning configurations of $\chi_x$ which differ by an integer multiple of $2\pi$ are identified. With this in mind, the final effective theory in Eq.~\eqref{eq:final effective theory} has a $\Z_2 \times \Z_2$ global symmetry structure generated by
\begin{equation}
    C: \chi_x \to -\chi_x+\pi \, \qquad S: \chi_x \to \chi_x+\pi \ .
\end{equation}
In particular, $C^2 = S^2 = 1$ and $CS = SC: \chi_x \to -\chi_x$. The global $\Z_2 \times \Z_2$ symmetry structure of this model is the same as that of the original height variables, although it is unclear how to trace the action of the original symmetries onto the scalar field in the effective theory.

The double cosine potential has minima at $\chi = \pm \pi/2$, which implies that the $C$ symmetry remains unbroken, while $S$ (and therefore also $CS$) is broken. This is again the same symmetry breaking structure that we observe numerically, as shown in Section \ref{sec:numerical}. We therefore identify the $C$ symmetry of the effective theory with charge conjugation and the $S$ symmetry with the single-site shift in the original model. While for the standard effective theory in Eq.~\eqref{eq:usual u(1) effective theory 2} all minima of the potential are physically equivalent because they are related by a global $\Z$ redundancy, this is no longer the case in our staggered model, where we find a proper $\Z_2$ symmetry breaking. 

Much like in the standard theory (see Section \ref{sec:usual u(1)}), the effective theory suggests two different ways of taking the continuum limit. In one case, one defines a mass (restoring the lattice spacing $a$)
\begin{equation}
    \label{eq:mass staggered prediction}
    a^2 m^2 = \frac{32\pi^4}{3} \frac{1}{e^4} \exp{\pqty{-4\pi^2 v_0 / e^2}} \ ,
\end{equation}
where $e^2$ is the dimensionless coupling. Similarly to the discussion in Section \ref{sec:usual u(1)}, keeping the mass $m$ fixed in physical units and sending the lattice spacing to zero forces $e^2 \to 0$. In this case, the height of the potential which separates the two physical minima becomes infinite, and therefore one finds that in the continuum this system consists of a free massive scalar field of mass $m$ describing the fluctuations around a spontaneously selected vacuum. In particular, Eq.~\eqref{eq:mass staggered prediction} is consistent with the qualitative numerical observation (see Fig.~\ref{fig:mass numerical results} and the discussion in Section \ref{sec:numerical}) that the exponent of the exponential decay of the mass is roughly twice as large in the staggered theory compared to the standard case.

It is also possible to take a different continuum limit, where the string tension instead is held fixed. In a semiclassical approximation \cite{Muller1981}, one again finds that also in this case the string tension scales like $\sigma \sim m e^2$. In the dual theory, the string tension can be interpreted as the interface tension between different minima of the potential. Since the staggered model has physically inequivalent minima, this continuum limit corresponds to a non-trivial theory where it is still possible to tunnel between the two inequivalent minima or to have domains of opposite minima separated by domain walls. The picture of the string tension as the interface tension between different broken symmetry ground states is also clear from Fig.~\ref{fig:local string structure}.

\section{Conclusions} \label{sec:conclusions}

In the present work, we have shown that the usual $\U(1)$ lattice gauge theory may be extended by an additional parameter $\extParam$ which modifies the Hilbert space per gauge link while preserving both the gauge symmetry and the remnant lattice Lorentz symmetry. We focused on the specific case of three spacetime dimensions with $\extParam=\pi$, which preserves all symmetries
of the standard $\U(1)$ theory, including charge conjugation and parity. After dualization, one obtains a height model which we have numerically simulated. We have also analytically derived an effective theory which provides predictions for the symmetry breaking structure as well as the scaling of the mass and string tension.  We have compared our numerical results for the staggered theory ($\extParam=\pi$) with our analytical predictions as well as with the analytical and numerical predictions available for the standard theory ($\extParam=0$). Our results provide evidence that the $\extParam=\pi$ theory has a spontaneously broken $\Z_2$ single-site shift symmetry, which remains unbroken in the standard theory, indicating that the $\extParam=\pi$ theory approaches a different continuum limit than the standard $\U(1)$ theory.

We have also obtained results for the mass and string tension. In particular, both quantities scale differently in the continuum limit compared to the ordinary theory. By inserting the charges directly in the partition function, we have also obtained data for the local energy distribution of the system with static charges, which shows that the confining flux string fractionalizes into two strands. Our numerical results are compatible with the analytical predictions from the effective theory for the $\extParam=\pi$ theory. The strands separate the system into two regions which live in different ground states of the broken $\Z_2$ symmetry.

\hyphenation{mani-fold} %
The idea behind this work may be extended in several directions. It would be immediate to generalize these results to a different number of spacetime dimensions. The Hamiltonian construction is valid in any dimension, but the dual theory would no longer be a scalar theory. For example, in four dimensions, the dual theory would be a gauge theory with a discrete gauge field. One could also study the theory at values of $\extParam \not\in \{0, \pi\}$, which explicitly breaks charge conjugation and parity, but may nonetheless be interesting gauge theories.
Perhaps the most interesting direction would be the extension to the non-Abelian case. Just as the introduction of an $\extParam$ angle in the $\U(1)$ theory is equivalent to threading a fictitious magnetic flux through the $\U(1)$ circle defining the Hilbert space at each link, an analogous construction is possible in the non-Abelian case as given in \cite{Chandrasekharan_2008, Vlasii:2018rwr}. In the case of $\SU(2)$, for example, this construction places a magnetic monopole at the center of the $3$-sphere manifold of $\SU(2)$, resulting in a modification of the Hilbert space of integrable functions on $\SU(2)$ while preserving the gauge symmetry.

\section*{Acknowledgments}

AB acknowledges financial support from the DAE, India. AB and DB acknowledge the computing resources of SINP. 
DB acknowledges assistance from SERB Starting grant SRG/2021/000396-C from the DST (Govt. of India).
GK, AM and UJW acknowledge funding from the Schweizerischer Nationalfonds through grant agreement no.~200020\_200424.
UJW also acknowledges support from the Alexander von Humboldt Foundation and thanks Ulf Meissner for hospitality at the Helmholtz-Institut f\"{u}r Strahlen-und-Kernphysik in Bonn. TR is supported by the Schweizerischer Nationalfonds through the grant no.~TMPFP2\_210064. Some calculations were performed on UBELIX (\url{http://www.id.unibe.ch/hpc}), the HPC cluster at the University of Bern.

\appendix

\section{Dualization to a height model in 3D} \label{sec:dualization}
The path integral of the 3D $\U(1)$ gauge theory can be dualized to a model of discrete height variables for any value of $\extParam$. The starting point for this dualization is given by the partition function in Eq.~\eqref{eq:extended partition fn}, reproduced below for convenience:
\begin{multline}
    \extPartFn \ .
\end{multline}
It is helpful to use Poisson summation to first rewrite this expression as
\begin{equation}
    Z = \left( \prod_l \int_0^{2\pi} d\varphi_l \right)
    \prod_p \left[
    \sum_{n \in \mathbb{Z}} e^{i (d\varphi)_p n} e^{-\frac{e^2}{2} (n + \frac{\extParam}{2\pi})^2}
    \right] \ .
\end{equation}
Note that one could also arrive at the above form by skipping the Poisson summation originally applied to derive Eq.~\eqref{eq:electric matrix elements}.

To render the integrals over the continuous $\U(1)$ variables tractable, we next rewrite the path integral over link variables into a path integral over plaquette variables,
\begin{equation}
    \left( \prod_l \int_0^{2\pi} d\varphi_l \right) =
    \left( \prod_p \int_0^{2\pi} d\Phi_p \right) \prod_c \left[ \sum_{h \in \mathbb{Z}} e^{i h (d\Phi)_c} \right] \ .
\end{equation}
Here, the product $\prod_c$ enumerates the three-dimensional unit cubes of the lattice, the Fourier series of the Dirac delta has been introduced using sums over $h$, and
\begin{equation}
    (d\Phi)_c = \sum_{i < j} \left[ \eps_{ijk} \Phi_{ij,x+\hat{k}} - \Phi_{ij,x} \right]
\end{equation}
is the exterior derivative of the plaquette variables $\Phi$ associated with cube $c$ rooted at site $x$. With this rewriting, the variables $(d\varphi)_p$ are replaced with the new integration variables $\Phi_p$, yielding
\begin{multline}
    Z = \left( \prod_c \sum_{h_c \in \mathbb{Z}} \right) \prod_p \Big[ \int_0^{2\pi} d\Phi_p \times \\
    \times \sum_{n \in \mathbb{Z}} e^{i \Phi_p (n - (d^*h)_p)}
    e^{-\frac{e^2}{2} (n + \frac{\extParam}{2\pi})^2}
    \Big] \ ,
\end{multline}
where $(d^* h)_p = \eps_{ijk} h_{x+\hat{k}} - h_x$ in terms of the orientation $(ij)$ and base site $x$ of the plaquette $p$.
Integrating out the $\Phi_p$ gives simple Kronecker delta functions over the integer sums, and resolving these results in the partition function of a height model,
\begin{equation}
    Z = \left( \prod_c \sum_{h_c \in \mathbb{Z}} \right) \prod_p \exp\pqty{-\frac{e^2}{2} ((d^*h)_p + \extParam / 2\pi)^2} \ .
\end{equation}

To complete the dualization, we go to the dual lattice. Cubes $c$ in the original lattice are dual to sites $x$ in the dual lattice, while plaquettes $p$ are dual to lattice edges $\left< xy \right>$, giving
\begin{equation}
    Z = \left( \prod_x \sum_{h_x \in \mathbb{Z}} \right) \exp\pqty{-\frac{e^2}{2} \sum_{\left< xy \right>} (h_x - h_y + \extParam / 2\pi)^2} \ ,
\end{equation}
where now all quantities refer to the dual lattice. Note that there is an ambiguity in the direction of the links $\left< x y \right>$ in the sum. Reversing the direction in which the link $\left< xy \right>$ appears in the sum is equivalent to locally flipping the sign of $\extParam$ on this link. One simple prescription, consistent with the largest translational symmetry group, is to choose the direction of the links in the sum to always be oriented in the forward $\hat{\mu}$ direction for each lattice orientation, meaning we sum over all links $\left< xy \right>$ with $y = x + \hat{\mu}$ for some $\mu \in \{0, 1, 2\}$. However, this prescription explicitly breaks lattice rotational symmetry, requiring each rotation to be combined with spatial inversions to remain a good symmetry of the theory. Instead, we choose to orient links $\left< xy \right>$ in the sum such that $x$ is always an odd site (i.e., $x_0 + x_1 + x_2$ is odd) and $y$ is therefore always an even site. This preserves lattice rotational symmetry and all translations by multiples of two lattice sites. The sum over odd sites can then be modified to absorb the $\extParam / 2\pi$ factor by instead summing over $h_x \in \mathbb{Z} + \extParam/2\pi$, resulting in the final dual partition function
\begin{multline}
    Z = \pqty{\prod_{x\, \mathrm{odd}} \sum_{h_x \in \Z + \frac{\extParam}{2 \pi}}} \pqty{\prod_{y\, \mathrm{even}} \sum_{h_y \in \Z}} \times \\ \times \exp{\bqty{-\frac{e^2}{2} \sum_{\expval{xy}} (h_x-h_y)^2}} \ .
\end{multline}
In this case, single-site translations are explicitly broken and must be combined with charge conjugation and a non-integer shift of the height variables to recover a symmetry transformation, as explained in Section \ref{sec:symmetries}. Note that for $\extParam \in \{0, \pi\}$ these two prescriptions are equivalent in infinite volume.

\section{Derivation of the energy operator}\label{sec:charge insertion energy derivation}

In order to measure the string tension, we insert static charges directly in the action and measure the resulting energy, similarly to  \cite{sterling-greensite, QuantumLink}. In particular, we know that the partition function $Z[s]$ with the charges inserted may be written in terms of a Hamiltonian $H$ as
\begin{equation}
    Z[s] = \tr{\pqty{e^{-\epsilon T H}}} \ ,
\end{equation}
where $\epsilon$ is the time step and $T$ is the number of time steps. 
Then we see that the energy of the system, i.e.\ the expectation value of the Hamiltonian, is given by
\begin{equation}
    \expval{H} = \frac{1}{Z}\tr{\pqty{H \, e^{-\epsilon T H}}} =-\frac{1}{Z T} \pdv{}{\epsilon} Z \ .
\end{equation}
Expressing the partition function in terms of the height variables and taking the appropriate derivative, one can then obtain an expression for $\expval{H}$ which is amenable to Monte Carlo simulation.

As we have seen in Section \ref{sec:string tension}, the partition function with the insertion of the two charges is given by
\begin{equation}
        Z = \pqty{\prod_{x \in \mathrm{sites}} \sum_{h_x}} \exp{\bqty{-\frac{e^2}{2} \sum_{\expval{xy}} (h_x-h_y+s_{\expval{xy}})^2}} \ ,
\end{equation}
where $s_{\expval{xy}}$ is defined in Eq.~\eqref{eq:sl}. In order to take the derivative, we restore the lattice spacing, which we set equal to $\epsilon$ in the time direction and equal to $a$ in the space directions. The sum over links discretizes an integral over volume, and is therefore associated with units of $\epsilon a^2$. The finite difference $h_x-h_y$ discretizes a derivative, and therefore comes with units of $1/a$ in the space directions and $1/\epsilon$ in the time direction; hence time links carry units of $a^2 \epsilon \times (1/\epsilon)^2=a^2/\epsilon$, while space links carry units of $a^2 \epsilon \times (1/a)^2=\epsilon$, so that overall the action becomes
\begin{equation}
\begin{aligned}
    S &= \frac{a^2 e^2}{2 \epsilon} \sum_{\expval{xy}_{\mathrm{time}}} (h_x-h_y+s_{\expval{xy}})^2 \\
    &\quad +\frac{\epsilon e^2}{2} \sum_{\expval{xy}_{\mathrm{space}}} (h_x-h_y+s_{\expval{xy}})^2 \ ,
\end{aligned}
\end{equation}
where $\expval{xy}_{\mathrm{time}}$ and $\expval{xy}_{\mathrm{space}}$ denote links in the time and space directions respectively. It is then straightforward to differentiate the partition function $Z[s]$ with respect to $\epsilon$. Setting back $\epsilon=a$, one then obtains (in lattice units)
\begin{equation} \label{eq:energy monte carlo}
\begin{aligned}
    \expval{H} &= \frac{e^2}{2T} \bigg\langle -\sum_{\expval{xy}_{\mathrm{time}}} (h_x-h_y+s_{\expval{xy}})^2 \\
    &\quad +\sum_{\expval{xy}_{\mathrm{space}}} (h_x-h_y+s_{\expval{xy}})^2\bigg\rangle \ .
\end{aligned}
\end{equation}
This expression can be computed via Monte Carlo simulation. It is important to note that, since we have a finite time step $\epsilon=a$, the expression $Z[s] = \tr\pqty{\exp{\pqty{-\epsilon T H}}}$ is valid only up to Trotterization artefacts. Therefore, we expect the equality in Eq.~\eqref{eq:energy monte carlo} to be valid only in the continuum limit $a \to 0$.

\bibliography{biblio}

\end{document}